\documentclass[
aps,prd,
12pt,
nopreprintnumbers,
showpacs,
eqsecnum,
nofootinbib
]{revtex4-1}

\usepackage{graphicx}
\usepackage{amssymb}

\newcommand{\be}{\begin{equation}}
\newcommand{\ee}{\end{equation}}
\newcommand{\bea}{\begin{eqnarray}}
\newcommand{\eea}{\end{eqnarray}}
\newcommand{\nn}{\nonumber \\}
\begin{document}

\preprint{AJC-HEP-25}
\title{
Evaluation~of vacuum energy density~around geometrical defects~by the
WKB method}
\author{Kiyoshi~Shiraishi
}
\address{Akita Junior College\\
Shimokitade-sakura, Akita-shi, Akita 010, Japan
}


\begin{abstract}
The energy density of a conformally invariant scalar field
around a cosmic string is estimated by the WKB method.
This approach reproduces the precise value obtained
 by the other authors using the exact mode-sum method or
the method of mirror images.
The approximation is also in good agreement with the exact result
in the three-dimensional case.
We also evaluate the energy density
of a conformally invariant scalar field around a global
monopole by the same way find the known result again.
\end{abstract}
\vspace{7mm}

\maketitle

\section{Introduction}
The vacuum energy of free fields on the conical space created by the presence of
an idealized cosmic string \cite{V} can be computed by the several
 ways \cite{qfcs}.
The metric describing an infinitely long straight cosmic string laid 
along the $z$ axis is given by \cite{V}
\be
ds^{2}=-dt^{2}+dz^{2}+dr^{2}+b^{2}r^{2}d\phi^{2},
\label{eq:met}
\ee
where $b$ is a parameter that characterizes the cosmic string.
Using this, the deficit angle is expressed as $2\pi(1-b)$
and the mass density of the string is $(1-b)/(4G)$.
For a conformally invariant scalar field, the vacuum expectation value
of the energy density at one-loop level
in the spacetime described by the metric (\ref{eq:met})
has been found to be \cite{qfcs}
\be
\rho_{vac} = -\frac{1}{1440\pi^{2}r^{4}}\left(\frac{1}{b^{4}}-1\right)\ .
\ee

In this paper, we evaluate the quantum vacuum energy of a conformally
invariant scalar field using the WKB method.
The derivation is simple and pedagogical:
Moreover, this yields considerably accurate values in many cases.
 As a quantum field,
 we consider only a conformally invariant scalar field in this paper.

The organization of this paper is as follows.
In Sec.~\ref{sec:cs} we study the WKB estimation of vacuum energy density
of a conformally invariant scalar field around a straight cosmic string.
The result is compared with the known one obtained by other methods.
 The technique is generalized to the cases of general spacetime
 dimensions in Sec.~\ref{sec:cone}\@.
 In Sec.~\ref{sec:gm},
we treat vacuum energy around a global monopole.
Sec,~\ref{sec:sum} is devoted to summary of the results.

\section{WKB evaluation of vacuum energy around a cosmic string}
\label{sec:cs}
The vacuum energy at one-loop level arises from the zero-point energy
 of quantum fields \cite{cas}.
Formally, it is written by
\be
E_{0}=\frac{1}{2}\sum_{\lambda}\omega_{\lambda}\ ,
\label{1}
\ee
where $\omega$ is the frequency of each normal mode
in the expansion of the field.
The subscript $\lambda$ characterizes each normal mode.
The formal expression (\ref{1}) diverges due to
unlimitedly high frequency modes.
Therefore, to get a meaningful conclusion for the vacuum energy
$E_{vac}$, we must regularize $E_{0}$
and subtract proper energy of standard,
which is usually defined for a flat Minkowsky space.

We start with the expression (\ref{1}) to obtain
vacuum energy density in the spacetime described
by the metric (\ref{eq:met})
representing the presence of a cosmic string.
We demonstrate evaluation of the vacuum energy
that comes from a conformally invariant scalar field
governed by the Lagrangian
\be
{\cal L}=\frac{1}{2}\nabla^{\mu}\varphi\nabla_{\mu}\varphi
+\frac{1}{2}\xi R \varphi^2,
\label{eq:lag}
\ee
where $R$ is the scalar curvature and
$\xi=1/6$ in the four-dimensional spacetime.

If the expression (\ref{1}) is calculated naively,
it is found to be divergent not only
because of the unlimitedly high frequency modes
but also because of the singularity of the space for the present case.
On the dimensional ground, the energy density should be proportional to
 $1/r^{4}$.
Other length or mass scales are absent in
the conformally invariant theory, at least in one-loop calculations.
Thus our aim can be said to be determination of the coefficient of $1/r^4$ in the
expression of the vacuum energy density.

The equation of motion for the scalar field is derived from
 (\ref{eq:lag}) as
\be
(\Box - {\xi} R)\varphi=0\ .
\label{eq:emo}
\ee
Now we define the following mode function:
\be
\varphi=e^{-i\omega t}e^{ikz}e^{i\ell \phi}\chi_{\omega k \ell}(r)\ ,
\label{eq:mode}
\ee
where $\ell=0, \pm 1, \pm 2,\ldots$.
Substituting the metric (\ref{eq:met}) and the
mode function (\ref{eq:mode}) into the wave equation (\ref{eq:emo}),
we get a differential equation for the radial function
 $\chi_{\omega k\ell}(r)$,
\be
\frac{1}{r}\frac{\partial}{\partial r}r\frac{\partial}{\partial r}
\chi_{\omega k\ell}(r)+
\left(\omega^{2}-k^{2}-\frac{\ell^{2}}{b^{2}r^{2}}\right)
\chi_{\omega k\ell}(r)=0\ .
\ee

To simplify the equation, we use the new coordinate $y$ defined by
\be
y=\ln{r/r_{0}}\ ,
\ee
where $r_{0}$ is an arbitrary constant.
Then the differential equation concerning the radial function becomes
\be
\frac{d^{2}\chi_{\omega k\ell}(y)}{dy^{2}}
+W(y;\omega,k,\ell)\chi_{\omega k\ell}(y)=0\ ,
\ee
where
\be
W(y;\omega,k,\ell)\equiv
(\omega^{2}-k^{2})r_{0}^{2}e^{2y}-\frac{\ell^{2}}{b^{2}}\ .
\ee

The solution according to the WKB approximation is given by
\be
\chi_{\omega k\ell}(y)\approx\frac{1}{\sqrt[4]{W(y;\omega,k,\ell)}}
\sin \int^{y}\sqrt{W(y;\omega,k,\ell)}\ dy\ .
\ee

Suppose that the ``wall'' is located at $r=L$,
which is an infrared cutoff in some sense.
On the other hand, near the conical singularity, we set a small cutoff
at $r=\delta$.
We assume the boundary conditions
$\chi_{\omega k\ell}=0$ at $r=\delta$ and $r=L$.
In this situation, a quantum number $n$ is associated with the radial
wave function as
\bea
n\pi&=&\int^{\ln L/r_{0}}_{\ln \delta/r_{0}}
\sqrt{W(y;\omega,k,\ell)}\ dy\nn
&=&\int^{L}_{\delta}\sqrt{\omega^{2}-k^{2}
-\frac{\ell^{2}}{b^{2}r^{2}}}\ dr\ .
\eea
Therefore the total number $g(\omega)$ of wave modes per unit length
along the cosmic string
with the frequency~$< \omega$ is written by
\be
g(\omega)=\int\frac{dk}{2\pi}
\sum^{}_{\ell}n(\omega,k,\ell)\ ,
\ee
where the sum over $\ell$ and integration over $k$ is taken just for real positive values of
the integrand.

We now get the vacuum energy per unit length along the cosmic string,
\bea
E_{0}&=&\frac{1}{2}\int dg(\omega)\ \omega \nn
&=&-\frac{1}{2}\int d\omega\ g(\omega) \nn
&=&-\frac{1}{2\pi}\int d\omega \int\frac{dk}{2\pi}
\sum^{}_{\ell}
 \int^{L}_{\delta}\sqrt{\omega^{2}-k^{2}
 -\frac{\ell^{2}}{b^{2}r^{2}}}\ dr\ .
\eea
Changing the order of integration, we have
\be
E_{0}=\int^{L}_{\delta}\ \rho_{0}(b)\ 2\pi
b\,rdr\ ,
\label{eq:ene}
\ee
with
\be
\rho_{0}(b)\equiv
-\frac{1}{4{\pi}^{2}br}\int d\omega \int\frac{dk}{2\pi}
\sum^{}_{\ell}
\sqrt{\omega^{2}-k^{2}-\frac{\ell^{2}}{b^{2}r^{2}}}\ .
\ee
The integration over $r$ in (\ref{eq:ene}) diverges
in the limit $\delta\rightarrow 0$,
since ``vacuum energy density'' $\rho_{0}(b)$ should be
proportional to
 $1/r^{4}$. Accordingly, we have only to manage
 the ultraviolet divergence
 in the formal expression of $\rho_{0}(b)$.

We again change the order of integration and perform the integration
over $\omega$ first. We treat the divergent integration by 
 analytical continuation. That is, the square root in the expression
 of $\rho_{0}(b)$ is regarded as the $(\frac{1}{2}-\epsilon)$th power.
The outcome of the integration is
\be
\rho_{0}(b)=-\frac{\mu^{2\epsilon}}{4\pi^{2}br}\int \frac{dk}{2\pi}
\sum^{\infty}_{\ell=-\infty}\frac{1}{\sqrt{4\pi}}
\Gamma\left(\frac{3}{2}-\epsilon\right)
\Gamma\left(-1+\epsilon\right)
\left(k^{2}+\frac{\ell^{2}}{b^{2}r^{2}}\right)^{1-\epsilon}\ ,
\ee
where the constant $\mu$ has the dimension of mass.

Next, we perform the integration over $k$. Then we get
\be
\rho_{0}(b)=-\frac{\mu^{2\epsilon}}{4\pi^{2}br}
\sum^{\infty}_{\ell=-\infty}\frac{1}{4\pi}
\Gamma\left(\frac{3}{2}-\epsilon\right)
\Gamma\left(-\frac{3}{2}+\epsilon \right)
\left(\frac{\ell^{2}}{b^{2}r^{2}}\right)^{\frac{3}{2}-\epsilon}.
\label{eq:unsum}
\ee
Discarding the term for $\ell=0$, we can rewrite the above as
\be
\rho_{0}(b)=-\frac{(\mu br)^{2\epsilon}}{8\pi^{3}(br)^{4}}
\Gamma\left(\frac{3}{2}-\epsilon\right)
\Gamma\left(-\frac{3}{2}+\epsilon \right)
\zeta\left(-3+2\epsilon\right)\ ,
\ee
where $\zeta(z)$ is the Riemann's zeta function.

Taking the limit $\epsilon\rightarrow 0$ in the context of analytic continuation,
we find the following finite quantity:
\bea
\rho_{0}(b)&=&-\frac{1}{8\pi^{3}(br)^{4}}
\Gamma\left(\frac{3}{2}\right)
\Gamma\left(-\frac{3}{2}\right)
\zeta\left(-3\right)\nn
&=&-\frac{1}{1440\pi^{2}(br)^{4}}\ ,
\eea
where the numerical values are substituted (see, for example,
\cite{AS}.).

Finally, the finite vacuum energy density is obtained by subtraction as
\bea
\rho_{vac}&=&\rho_{0}(b)-\rho_{0}(1)\nn
&=&-\frac{1}{1440\pi^{2}r^{4}}\left(\frac{1}{b^{4}}-1\right).
\eea
This result coincides with the known result obtained
by exact mode summation or mirror-image method \cite{qfcs}.

We can also consider a ``twisted'' scalar field \cite{twist}, which has a
special  periodicity in $\phi$: 
$\varphi(t,z,r,\phi)=-\varphi(t,z,r,\phi+2\pi)$.
For the twiwted field, the vacuum energy can be obtained by replacing
$\ell\rightarrow \ell+1/2$.
Thus the vacuum energy density before subtraction is derived as
\be
\rho_{0}^{T}(b)=-\frac{7}{8}\rho_{0}(b)\ .
\ee
Then the finite vacuum energy density is given by
\bea
\rho_{vac}^{T}&=&\rho_{0}^{T}(b)-\rho_{0}(1)\nn
&=&\frac{7b^{-4}+8}{11520\pi^{2}r^{4}}\ ,
\eea
which also turns out to be the exact expression \cite{qfcs}.

In the next section, the procedure of evaluation of vacuum energy
 is generalized to an arbitrary dimensional case.

\section{Vacuum energy for a conical defect in general dimensions}
\label{sec:cone}
In this section, we consider a conical singularity
 in general dimensions.
The $(d+3)$-dimensional metric is written by
\be
ds^{2}=-dt^{2}+(dx^{1})^{2}+\cdots +(dx^{d})^{2}
+dr^{2}+b^{2}r^{2}d\phi^{2}.
\label{eq:metg}
\ee
The Lagrangian for a conformal scalar field is the same as (\ref{eq:lag}),
except for $\xi=(d+1)/[4(d+2)]$.

The evaluation of vacuum energy density of a conformal scalar field
in this spacetime is done by a similar way that we showed
in the preceding section.
In our methods, the unregularized energy density per unit $d$-volume can be expressed as
\be
\rho_{0}(b)=-\frac{\mu^{2\epsilon}}{4\pi^{2}br}\int \frac{d^{d}k}{(2\pi)^{d}}
\sum^{\infty}_{\ell=-\infty}\frac{1}{\sqrt{4\pi}}
\Gamma\left(\frac{3}{2}-\epsilon\right)
\Gamma\left(-1+\epsilon \right)
\left(\sum^{d}_{i=1}(k^{i})^{2}+\frac{\ell^{2}}{b^{2}r^{2}}
\right)^{1-\epsilon}.
\ee
The difference from the previous section is the dimension of the integration over $k$'s.
Carrying out the integration over $k^{i}\ (i=1,\ldots,d)$ and arranging
the sum over $\ell$,
we have
\be
\rho_{0}(b)=-\frac{(\mu br)^{2\epsilon}}{2\pi^{2}
(4\pi)^{(d+1)/2}(br)^{d+3}}
\Gamma\left(\frac{3}{2}-\epsilon\right)
\Gamma\left(-\frac{d}{2}-1+\epsilon\right)
\zeta\left(-d-2+2\epsilon\right).
\label{rd}
\ee
Further applying the reciprocal formula for zeta and gamma functions
\cite{Er}
\be
\zeta(z)\Gamma\left(\frac{z}{2}\right)=
\pi^{z-1/2}\zeta(1-z)\Gamma\left(\frac{1-z}{2}\right)
\ee
to (\ref{rd}), we obtain
\be
\rho_{0}(b)=-\frac{1}{2^{d+3}\pi^{3(d+3)/2}(br)^{d+3}}
\Gamma\left(\frac{d+3}{2}\right)
\zeta\left(d+3\right),
\ee
where the limit $\epsilon\rightarrow 0$ has been taken.


Consequently, the regularized vacuum energy density
for an untwisted conformal scalar filed is given by
\bea
\rho_{vac}&=&\rho_{0}(b)-\rho_{0}(1)\nn
&=&-\frac{1}{(4\pi^{3})^{(d+3)/2}r^{d+3}}
\Gamma\left(\frac{d+3}{2}\right)
\zeta\left(d+3\right)
\left(\frac{1}{b^{d+3}}-1\right)
\label{eq:ved}
\eea

An exact result for $d=0$ has been shown by Souradeep and Sahni \cite{SS}.
 Their result is given by the form of integration,
\be
\rho_{vac:exact}=-\frac{1}{16\pi^{2}r^{3}}
\int^{\infty}_{0}\frac{du}{\sinh  u}
\left(
\frac{\coth  u}{\sinh^{2}  u}-
\frac{\coth  u/b}{b^{3}\sinh^{2}  u/b}
\right),
\label{eq:SSresult}
\ee
where the notation has been changed into ours.

On the other hand, our approximation (\ref{eq:ved}) gives
\be
\rho_{vac:WKB}=-\frac{1}{16\pi^{4}r^{3}}
\zeta\left(3\right)
\left(\frac{1}{b^3}-1\right)\ ,
\ee
for $d=0$.

\begin{figure}
\centering
\includegraphics[width=8cm]
{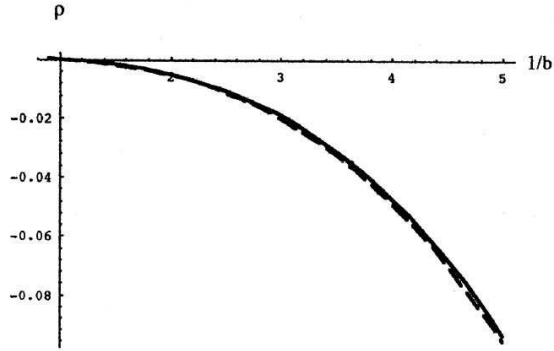}
\caption{$r^{3}\rho_{vac}$ as a function of $b^{-1}$.
The solid line represents the exact value
 $r^{3}\rho_{vac:exact}$, while the dashed line represents
 our approximation
 $r^{3}\rho_{vac:WKB}$.}
\label{fig1}
\end{figure}

The comparison between the exact and approximate results
is displayed in FIG. \ref{fig1}. We conclude that the result of the
WKB approximation is
in excellent agreement with the exact value for the three-dimensional case ($d=0$).

\section{Vacuum energy around a global monopole}
\label{sec:gm}
A global monopole \cite{BV}
can be regarded as a geometrical point singularity with
a deficit solid angle, if the finite-mass contribution is neglected.
The metric of the spacetime describing a global monopole is
\be
ds^{2}=-dt^{2}+dr^{2}+b^{2}r^{2}(d\theta^{2}
+\sin^{2}\theta\ d\phi^{2})\ .
\label{eq:1gm}
\ee

It is possible to generalize the metric to the ($d+N+1$)-dimensional one,
\be
ds^{2}=-dt^{2}+(dx^{1})^{2}+\cdots+(dx^{d})^{2}+
dr^{2}+b^{2}r^{2}d\Omega_{N-1}^{2}\  ,
\label{eq:metdn1}
\ee
where $d\Omega_{N-1}^{2}$ is a line element on an $(N-1)$-sphere
 with a unit radius.

The derivation of the formal expression for the vacuum energy of
a conformal scalar field in the spacetime described by the metric (\ref{eq:metdn1}) can be done
 similarly to that in the previous sections.

We have only to notice a few differences.
\begin{itemize}
\item The coupling to the scalar curvature $\xi$ is now
$(d+N-1)/[4(d+N)]$.
\item The value of the scalar curvature is nonzero for $N\geq 3$.
\item The mode function takes the form
\[
\varphi=
\frac{1}{r^{(N-2)/2}}
e^{-i\omega  t}e^{ikz}Y_{\ell\nu}^{(N-1)}(\Omega)
\chi_{\omega k \ell}(r)\ ,
\]
where
$Y_{\ell\nu}^{(N-1)}(\Omega)$ is the generalized spherical function
whose eigen value for the laplacian on $S^{N-1}$ is $\ell(\ell+N-2)$.
\end{itemize}

Hence we obtain
\be
\rho_{0}(b)=
-\frac{(\mu  br)^{2\epsilon}\Gamma(N/2)}{2^{d+3}\pi^{(d+N+3)/2}
(br)^{d+N+1}}
\Gamma\left(\frac{3}{2}-\epsilon\right)
\Gamma\left(-\frac{d}{2}-1+\epsilon\right)
\sum^{\infty}_{\ell=0}D_{\ell}
{\left[{\left(\ell+\frac{N-2}{2}\right)}^{2}+\sigma^{2}\right]}
^{\frac{d}{2}+1-\epsilon}\  ,
\label{eq:rbdn1}
\ee
where
\be
D_{\ell}\equiv
\frac{(2\ell+N-2)(\ell+N-3)!}{(N-2)!\ell !}\ ,
\ee
and
\be
\sigma^{2}\equiv
\frac{(d+1)(N-2)}{4(d+N)}(1-b^{2})\ .
\ee
The choice $d=1$ and $N=2$ leads to the vacuum energy around
 a cosmic string (\ref{eq:unsum}), etc..

Various ways to regularize the divergent quantity are
 known \cite{CW}.
 Instead of investigating general treatment, we study here the
 ``original'' case
with the metric (\ref{eq:1gm})
($d=0$ and $N=3$), i.e., the case with
a global monopole at the origin.
 The generalization of this example
to other cases is straightforward.

For $d=0$ and $N=3$, (\ref{eq:rbdn1}) becomes
\be
\rho_{0}(b)=-\frac{(\mu br)^{2\epsilon}}{2^{4}\pi^{5/2}(br)^{4}}
\Gamma\left(\frac{3}{2}-\epsilon\right)
\Gamma\left(-1+\epsilon\right)
\sum^{\infty}_{\ell=0}\ (2\ell+1)
{\left[{\left(\ell+\frac{1}{2}\right)}^{2}+\sigma^{2}\right]}
^{1-\epsilon}\ ,
\label{eq:rbgm}
\ee
with
\be
\sigma^{2}\equiv
\frac{1-b^{2}}{12}\ .
\ee
Using an expansion \cite{CW}
\be
{\left[{\left(\ell+\frac{1}{2}\right)}^{2}+\sigma^{2}\right]}
^{1-\epsilon}
=\sum^{\infty}_{m=0}\frac{\Gamma(m-1+\epsilon)}
{\Gamma(-1+\epsilon)\ m!} (-\sigma^{2})^{m}
{\left(\ell+\frac{1}{2}\right)}^{2-2\epsilon-2m}\ ,
\ee
we rewrite (\ref{eq:rbgm}) as
\be
\rho_{0}(b)=-\frac{(\mu br)^{2\epsilon}}{2^{3}\pi^{5/2}(br)^{4}}
\Gamma\left(\frac{3}{2}-\epsilon\right)
\sum^{\infty}_{m=0}\frac{\Gamma(m-1+\epsilon)}
{m!} (-\sigma^{2})^{m}
(2^{2m-3+2\epsilon}-1)\zeta(2m-3+2\epsilon)\ .
\ee

For the first three terms in the sum diverges as $1/\epsilon$
in the limit $\epsilon\rightarrow 0$,
we introduce the renormalization scale $\bar{\mu}$,
which involves the mass parameter $\mu$.
We can regard that the divergence and other finite terms
are absorbed into the choice of the renormalization scale.

Then we get the following expression
in the limit of $\epsilon\rightarrow 0$:
\be
\rho_{0}(b)=
-\frac{1}{768\pi^{2}(br)^{4}}
\left[\frac{7}{10}-\frac{1}{3}(1-b^2)+\frac{1}{6}(1-b^2)^2
\right]
\ln \bar{\mu} r\ .
\label{eq:gg}
\ee

The result should be  compared with the one obtained
by Mazzitelli and Lousto \cite{ML}.
In our notation, their result in \cite{ML} should read as
\be
\rho_{vac}=-\frac{1}{720\pi^{2}(b\,r)^{4}}(1-b^{2})
\left(1-\frac{1-b^{2}}{2}\right)\ \ln \bar{\mu} r\ ,
\ee
since a factor two has been missed there (cf.\ \cite{BD}).

One can easily find that this coincides with the vacuum energy density
obtained by our WKB approach $\rho_{vac}=\rho_{0}(b)-\rho_{0}(1)$, where
$\rho_{0}(b)$ is given in (\ref{eq:gg}).

\section{Summary}
\label{sec:sum}

We have shown the validity of the WKB method to obtain
one-loop vacuum energy density of a conformally invariant scalar
field around geometrical defects.

In particular, for the case with
a straight casmic string in four dimensional spacetime,
we have found that the WKB evaluation reproduces
an exact value of the vacuum energy density
for untwisted or twisted conformal scalar fields.
In the case with a three-dimensional conical spacetime,
the WKB result is not an exact value
but an excellent approximation of the vacuum energy density for
a conformally invariant scalar field.
The vacuum energy density around a global monopole has been
calculated by the WKB method   for
a conformally invariant scalar field.
The result coincides with the one obtained by another method.

The varidity of our WKB approach
is ensured by the following.
\begin{itemize}
\item The background metric of the present model
 contains no dimensionful constants except for $r$.
\item The Lagrangian for a scalar field in the present model
 is conformally invariant.
\item The regularized quantity is considered to be largely relied
on the divergence from rather high frequency modes.
\end{itemize}

It will be worth studying generalization of the WKB method
 to evaluation of free energy density of
 quantum fields in finite-temperature system with
geometrical defects.

\bibliographystyle{apsrev4-1}

\end{document}